\documentclass{acm_proc_article-sp}
\pdfoutput=1
\usepackage{multirow}
\usepackage{color}
\usepackage{array}

\usepackage[bookmarks=true%
,bookmarksnumbered=true%
,hypertexnames=false%
,breaklinks=true%
,colorlinks=true%
,linkcolor=blue%
,citecolor=blue%
,urlcolor=blue%
]{hyperref}\hypersetup{
  pdfauthor = {},
  pdftitle = {Towards Real-Time Summarization of Scheduled Events from Twitter Streams},
  pdfsubject = {},
  pdfkeywords = {H.3 [Information Storage and Retrieval]},
  pdfcreator = {},
  pdfproducer = {}
}

\newcommand{\comment}[1]{}

\definecolor{Orange}{rgb}{1,0.5,0}

\begin{document}

\title{Towards Real-Time Summarization of Scheduled Events from Twitter Streams
\titlenote{The present paper gives more technical and experimental details about the work published as a
poster at HT'2012~\cite{zubiaga2012summarization}.}}
\numberofauthors{4} 

\author{
\alignauthor
Arkaitz Zubiaga\\
\affaddr{Queens College}\\
\affaddr{City University of New York}\\
\affaddr{New York, NY, USA}\\
\email{arkaitz.zubiaga@qc.cuny.edu}
\alignauthor
Damiano Spina\\
\affaddr{NLP\&IR Group} \\
\affaddr{ETSI Inform\'{a}tica UNED}\\
\affaddr{Madrid, Spain}\\
\email{\hspace{2ex}damiano@lsi.uned.es}
\and
\alignauthor
Enrique Amig\'{o}\\
\affaddr{NLP\&IR Group} \\
\affaddr{ETSI Inform\'{a}tica UNED}\\
\affaddr{Madrid, Spain}\\
\email{enrique@lsi.uned.es}
\alignauthor
Julio Gonzalo\\
\affaddr{NLP\&IR Group} \\
\affaddr{ETSI Inform\'{a}tica UNED}\\
\affaddr{Madrid, Spain}\\
\email{julio@lsi.uned.es}
}

\maketitle
\begin{abstract}
This paper explores the real-time summarization of scheduled events such as soccer games from
torrential flows of Twitter streams. We propose and evaluate an approach that
substantially shrinks
the stream of tweets in real-time, and consists of two steps: (i) sub-event detection, which
determines if something new has occurred, and (ii) tweet selection, which picks a representative
tweet to describe each sub-event. We compare the summaries generated in three languages for all the
soccer games in \emph{Copa America 2011} to reference live reports offered by Yahoo! Sports
journalists. We show that simple text analysis methods which do not involve external knowledge lead
to summaries that cover 84\% of the sub-events on average, and 100\% of key types of sub-events
(such as goals in soccer). Our approach should be straightforwardly applicable to other kinds of
scheduled events such as other sports, award ceremonies, keynote talks, TV shows, etc.
\end{abstract}




\section{Introduction}
\label{sec:introduction}
Twitter\footnote{http://twitter.com/} has gained widespread popularity as a
microblogging site where
users share short messages (\emph{tweets}). Twitter users not only tweet about
their personal issues
or nearby events, but also about news and events of interest to some
community~\cite{mishaud2007}.
Twitter has become a powerful tool to stay tuned to current affairs. It is known that, in
particular, Twitter users exhaustively share messages about (all kinds of) events they are following
live, occasionally giving rise to related trending
topics~\cite{zubiaga2011classifying}. 

The community of users live \emph{tweeting} about a given event generates rich
contents describing sub-events that occur during an event (e.g., goals, red
cards or penalties in a soccer game). All those users share valuable information
providing live coverage of events~\cite{becker2012identifying}. However, this
overwhelming amount of
information makes difficult for the user: (i) to follow the full stream while finding out about new
sub-events, and (ii) to retrieve from Twitter the main, summarized information about which are the
key things happening at the event. In the context of exploring the potential of
Twitter as a means
to follow an event, we address the (yet largely unexplored) task of summarizing Twitter contents by
providing the user with a summed up stream that describes the key sub-events. We propose a two-step
process for the real-time summarization of events --sub-event detection and tweet selection--, and
analyze and evaluate different approaches for each of these two steps. We find that Twitter provides
an outstanding means for detailed tracking of events, and present an approach that accurately
summarizes streams to help the user find out what is happening throughout an event. We perform
experiments on scheduled events, where the start time is known. By comparing different
summarization approaches, we find that learning from the information seen before throughout the
event is really helpful both to determine if a sub-event occurred, and to select a tweet that
represents it.

To the best of our knowledge, our work is the first to provide an approach to generate real-time
summaries of events from Twitter streams without making use of external knowledge. Thus, our
approach might be straightforwardly applied to other kinds of scheduled events without requiring
additional knowledge. 

\section{Dataset}
\label{sec:dataset}

We study the case of tweets sent during the games of a soccer competition.
Sports events are
a good choice to explore for summarization purposes, because they are usually reported live by
journalists, providing a reference to compare with. We set out to explore the \emph{Copa America
2011} championship, which
took place from July $1^{\text{st}}$ to $24^{\text{th}}$, 2011, in
Argentina, where 26 soccer games were played. Choosing an
international competition with a wide reach enables to gather and summarize tweets in different
languages. The official start times for the games were announced in advance by the organization.

During the period of the \emph{Copa America}, we gathered all the tweets that
contained any of \texttt{\#ca2011}, \texttt{\#copaamerica}, and
\texttt{\#copaamerica2011}, which were set to be the official Twitter hashtags
for the competition. For the 24 days of collection, we retrieved 1,425,858
unique tweets sent by 290,716 different users. These tweets are written in 30
different languages, with a majority of 76.2\% in Spanish, 7.8\% in Portuguese,
and 6.2\% in English. The tweeting activity of the games considerably varies,
from 11k tweets for the least-active game, to 74k for the most-active one, with
an average of 32k tweets per game.

In order to define a reference for evaluation, we collected the live reports for
all the games given by Yahoo!
Sports\footnote{\texttt{
http://uk.eurosport.yahoo.com/football/ \mbox{copa-america/fixtures-results/}}}.
These reports include the annotations of
the most relevant sub-events throughout a game. 7 types of annotations are included: goals (54 were
found for the 26 games), penalties (2), red cards (12), disallowed goals
(10), game starts (26),
ends (26), and stops and resumptions (63). On average, each game comprises 7.42 annotations. Each
of these annotations includes the minute when it happened. We manually annotated the beginning of
each game in the Twitter streams, so that we could infer the timestamp of each annotation from
those minutes. The annotations do not provide specific times with seconds, and the actual timestamp
may vary slightly. We have considered these differences for the evaluation process.

\section{Real-Time Event Summarization}
\label{sec:event-summarization}

\begin{figure}[hbt]
 \begin{center}
  \includegraphics[scale=0.85]{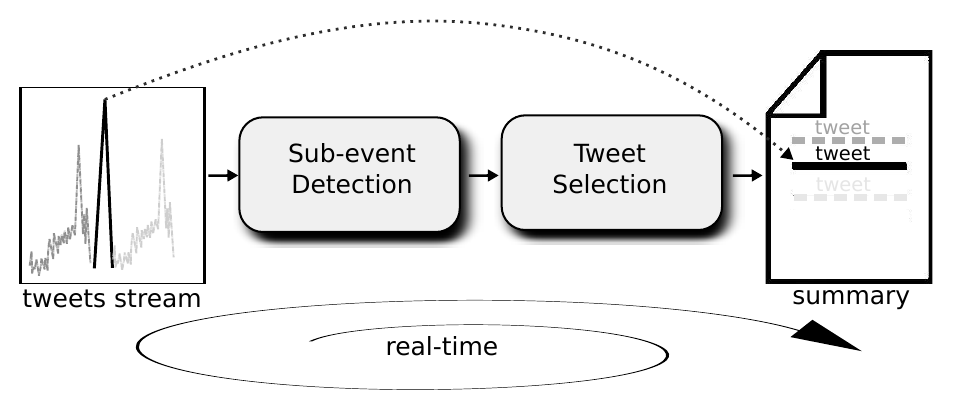}
  \caption{Two-step process for real-time event summarization.}
  \label{fig:architecture}
 \end{center}
\end{figure}

We define real-time event summarization as the task that provides new
information about an event every time a relevant sub-event occurs. To tackle the
summarization task, we define a two-step process that enables to report
information about new sub-events in different languages. The first step is to
identify at all times whether or not a specific sub-event occurred in the last few seconds.
The output will be a boolean value determining if something relevant occurred;
if so, the second step is to choose a representative tweet that describes the
sub-event in the language preferred by the user. The aggregation of these two
processes will in turn provide a set of tweets as a summary of the game (see
Figure~\ref{fig:architecture}).

\subsection{First Step: Sub-Event Detection}
\label{sec:event-detection}

\begin{figure}[hbt]
 \begin{center}
  \includegraphics[width=0.475\textwidth]{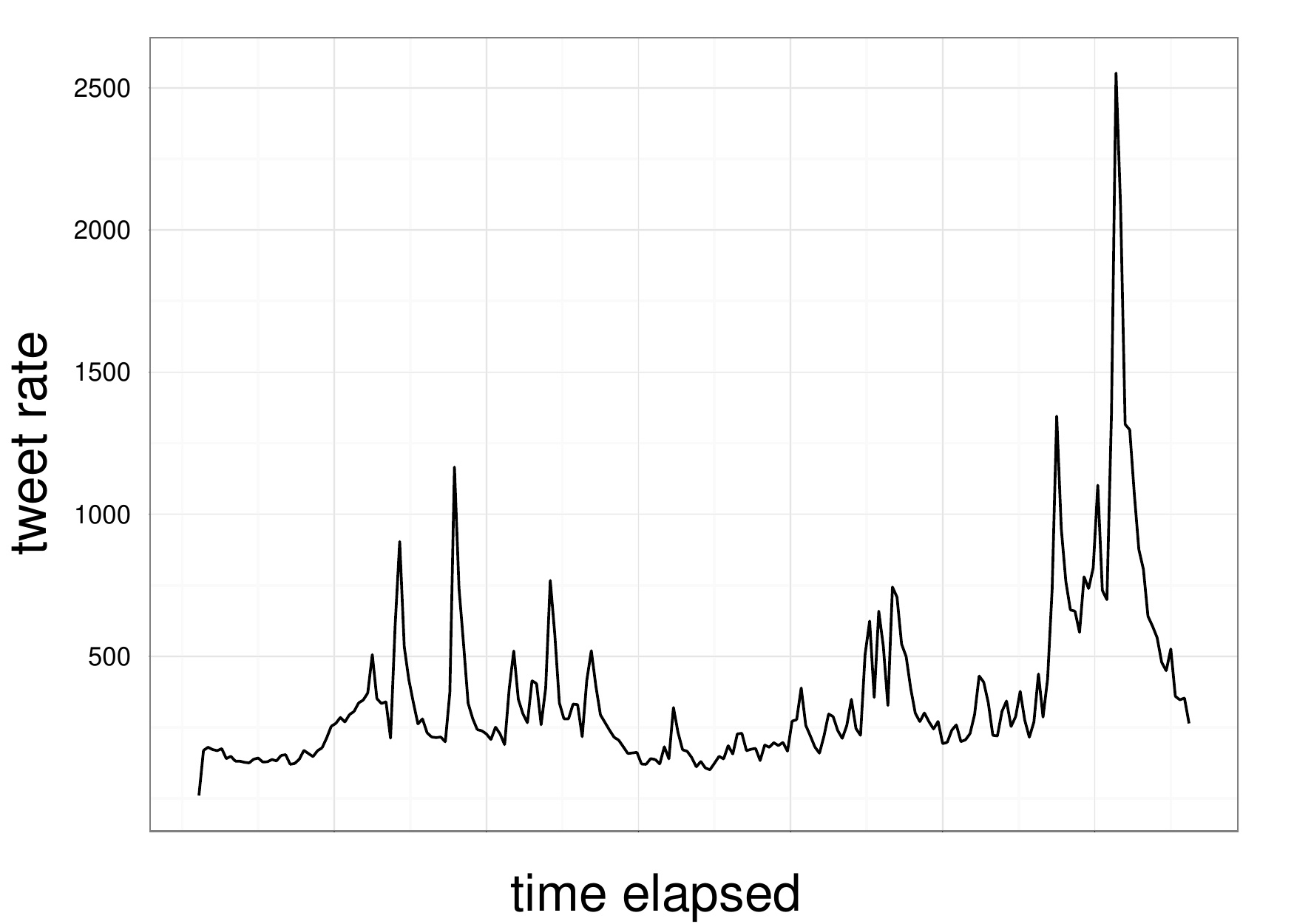}
  \caption{Sample histogram of tweeting rates for a soccer game (Argentina vs Uruguay), where
several peaks can be seen.}
  \label{fig:game-histogram}
 \end{center}
\end{figure}

The first part of the event summarization system corresponds to the sub-event detection. Note that,
being a real-time sub-event detection, the system has to determine at all times whether or not a
relevant sub-event has occurred, clueless of how the stream will continue to evolve. Before the
beginning of an event, the system is provided with the time that it starts, as scheduled in advance,
so the system knows when to start looking for new sub-events. With the goal of developing a
real-time sub-event detection method, we rely on the fact that relevant sub-events trigger a massive
tweeting activity of the community. We assume that the more important a sub-event is, the more users
will tweet about it almost immediately. This is reflected as peaks in the histogram of tweeting
rates (see Figure~\ref{fig:game-histogram} for an example of a game in our
dataset). In the
process of detecting sub-events, we aim to compare 2 different ideas: (i) considering only sudden
increase with respect to the recent tweeting activity, and (ii) considering also all the previous
activity seen during a game, so that the system learns from the evolution of the audience. We
compare the following two methods that rely on these 2 ideas:

\begin{enumerate}
 \item \textbf{Increase:} this approach was introduced by Zhao et
al.~\cite{zhao2011human}. It
considers that an important sub-event will be reflected as a sudden increase in the tweeting rate.
For time periods defined at 10, 20, 30 and 60 seconds, this method checks if the tweeting rate
increases by at least 1.7 from the previous time frame for any of those
periods. If the increase
actually occurred, it is considered that a sub-event occurred. A potential
drawback of this method is
that not only outstanding tweeting rates would be reported as sub-events, but also low rates that
are preceded by even lower rates.

 \item \textbf{Outliers:} we introduce an outlier-based approach that relies on whether the tweeting
rate for a given time frame stands out from the regular tweeting rate seen so far during the event
(not only from the previous time frame). We set the time period at 60 seconds for this approach. 15
minutes before the game starts, the system begins to learn from the tweeting rates, to find out what
is the approximate audience of the event. When the start time approaches, the system begins with the
sub-event detection process. The system considers that a sub-event occurred when the tweeting rate
represents an outlier as compared to the activity seen before.
Specifically, if the tweeting
rate is above 90\% of all the previously seen tweeting rates,
the current time frame will be
reported as a sub-event. This threshold has been set a priori and without
optimization. The outlier-based method incrementally learns while the game
advances, comparing the
current tweeting rate to all the rates seen previously. Different from the increase-based approach,
our method presents the advantages that it considers the specific audience of an event, and that
consecutive sub-events can also be detected if the tweeting rate remains constant without increase.
Accordingly, this method will not consider that a sub-event occurred for low tweeting rates
preceded by even lower rates, as opposed to the increase-based approach.
\end{enumerate}

Since the annotations on the reference are limited to minutes, we round down the
outputs of the systems to match the reference. Also, the timestamps annotated
for the reference are not entirely precise, and therefore we accept as a correct guess an automatic
sub-event detection that differs by at most one minute from the reference.

This evaluation method enables us to compare the two systems to infer which of them performs best.
Table~\ref{tab:baseline-outliers-evaluation} shows the precision (P), recall (R)
and F-measure (F1)
of the automatically detected sub-events with respect to the reference, as well as the average
number of sub-events detected per game (\#). Our outliers approach clearly outperforms the baseline,
improving both precision (75.8\% improvement) and recall (3.7\%) for an overall
40\% gain in F1. At
the same time, the compression rate for the outliers approach almost doubles that of the baseline
(56.4\%). From the average of 32k tweets sent per game, the summarization to 25.6 tweets represents
a drastic reduction to only 0.079\% of the total. Keeping the number of sub-events small while
effectiveness improves is important for a summarization system in order to provide a concise and
accurate summary. The outperformance of the outlier-based approach shows the importance of taking
into account the audience of a specific game, as well as the helpfulness of learning from previous
activity throughout a game.

\begin{table}[hbt]
 \begin{center}
  \begin{tabular}{| l | c | c | c | c |}
   \cline{2-5}
   \multicolumn{1}{ l |}{} & \textbf{P} & \textbf{R} & \textbf{F1} & \textbf{\#}
\\
   \hline
   \hline
   \textbf{Increase} & 0.29 & 0.81 & 0.41 & 45.4 \\
   \hline
   \textbf{Outliers} & \textbf{0.51} & \textbf{0.84} & \textbf{0.63} &
\textbf{25.6} \\
   \hline
  \end{tabular}
  \caption{Evaluation of sub-event detection approaches.}
  \label{tab:baseline-outliers-evaluation}
 \end{center}
\end{table}

\subsection{Second Step: Tweet Selection}
\label{sec:tweet-selection}

The second and final part of the summarization system is the tweet selection.
This second step is only activated when the first step reports that a new sub-event occurred.
Once the system has determined that a sub-event occurred, the tweet
selector is provided with the tweets corresponding to the minute of the
sub-event. From those tweets, the system has to choose one as a representative
tweet that describes what occurred. This tweet must provide the main information about the
sub-event, so the user understands what occurred and can follow the event. Here we compare two tweet
selection methods, one relying only on information contained within the minute of the sub-event,
and another considering the knowledge acquired during the game. We test them on the output of the
outlier-based sub-event detection approach described above, as the approach with best performance
for the first step.

To select a representative tweet, we get a ranking of all the tweets. To do so, we score each tweet
with the sum of the values of the terms that it contains. The more representative are the terms
contained in a tweet, the more representative will be the tweet itself. To define the values of the
terms, we compare two methods: (i) considering only the tweets within the sub-event (to give
highest values to terms that are used frequently within the sub-event), and (ii) taking into account
also the tweets sent before throughout the game, so that the system can make a difference from
what has been the common vocabulary during the event (to give highest values to
terms that are especially used within the minute and not so frequently earlier during the event).
We use the following well-known approaches to implement these two ideas:

\begin{enumerate}
 \item \textbf{TF:} each term is given the value of its frequency as the number of occurrences
within the minute, regardless of its prior use.
 \item \textbf{KLD:} we use the Kullback-Leibler divergence
\cite{kullback1951information} (see Equation \ref{eq:kld}) to measure how
frequent is a term $t$ within the sub-event ($H$), but also considering how
frequent it has been during the game until the previous minute ($G$). Thus, KLD
will give a higher weight to terms frequent within the minute that were less
frequent during the game. This may allow to get rid of the common vocabulary all along the game,
and rather provide higher rates to specific terms within the sub-event.
\end{enumerate}

\begin{equation}
 D_{\mathrm{KL}}(H\|G) = H(t) \log \frac{H(t)}{G(t)}
 \label{eq:kld}
\end{equation}

With these two approaches, the sum of values for terms contained in each tweet results in a weight
for each tweet. With weights given to all tweets, we create a ranking of tweets sent during the
sub-event, where the tweet with highest weight ranks first. We create these rankings for each of
the languages we are working on. The tweet that maximizes this score for a given language is
returned as the candidate tweet to show in the summary in that language. The two term weighting
methods were applied to create summaries in three different languages: Spanish, English, and
Portuguese. We test them on
the output of the outlier-based sub-event detection approach
described above, as the approach with best performance for
the first step. Thus, we got six summaries for each game, i.e., TF and KLD-based
summaries for the
three languages. These six summaries were manually evaluated by comparing them
to the reference. Table~\ref{tab:example-tweets} shows some tweets included
in the KLD-based summary in English.

\begin{table*}[htb]
 \begin{center}
  \begin{tabular}{ |
>{\centering\let\newline\\\arraybackslash\hspace{0pt}}m{2cm} |
>{\let\newline\\\arraybackslash\hspace{0pt}}m{7cm} |
>{\let\newline\\\arraybackslash\hspace{0pt}}m{7cm} | }
   \hline
  \centering \small \textbf{Sub-event} & \small \centering \textbf{Selected
Tweet} & \small \hspace{1.8cm}\textbf{Narrator's Comment} \\
   \hline
   \small Game start & \small RT
@user: Uruguay-Argentina. The R\'io de la Plata classic. The 4th vs the 5th
in the last WC. History doesn't matter. Argentina must win. \#ca2011 & \small
The referee gets the game under way \\
   \hline
\small Goal & \small Gol!
Gol! Gol! de Perez  Uruguay 1 vs Argentina 0  Such a quick strike and Uruguay is
already on top. \#copaamerica & \small GOAL!! Forlan's free kick is hit deep
into the box and is
flicked on by Caceres. Romero gets a hand on it but can only push it into the
path of Perez who calmly strokes the ball into the net. \\
\hline
\small Goal & \small Gooooooooooooooooal Argentina ! Amazing pass from Messi, Great positioning \&
finish from Higuain !! Arg 1 - 1 Uru \#CopaAmerica & \small GOAL!! Fantastic response from
Argentina. Messi picks the ball up on the right wing and cuts in past Caceres. The Barca man clips a
ball over the top of the defence towards Higuain who heads into the bottom corner. \\
\hline
\small Red card & \small Red card
for Diego P\'erez, his second yellow card, Uruguay is down to 10, I don't know
if I would have given it. \#CopaAm\'erica2011 & \small You could see it coming.
How stupid. Another needless
free kick conceded by Perez and this time he is given his marching order. He
purposely blocks off Gago. Uruguay have really got it all to do now. \\
\hline
\small Red card & \small \#ca2011 Yellow for Mascherano! Double yellow! Adios! 10 vs 10! Mascherano
surrenders his captain armband! & \small It's ten against ten. Macherano comes across and fouls
Suarez. He's given his second yellow and his subsequent red. \\
\hline
\small Game stop (full time) & \small Batista didn't look too happy at the
game going to
penalties as the TV cut to hit at FT, didn't appear confident \#CA2011 & \small
The second half is brought to an end. We will have
extra time. \\
\hline
\small Game end & \small Uruguay beats Argentina! 1-1 (5-4 penalty shoot out)!
Uruguay now takes on Peru in Semis. \#copaamerica & \small ARGENTINA 4-5 -
URUGUAY WIN. Caceres buries the final penalty into the top right-hand corner. \\
\hline
  \end{tabular}
 \end{center}
 \caption{Example of some tweets selected by the (outliers+KLD) summarization
system, compared with the respective comments narrated on Yahoo! Sports.}
 \label{tab:example-tweets}
\end{table*}

In the manual evaluation process, each tweet in a system summary is classified as correct if it can
be associated to a sub-event in the reference and is descriptive enough (note that there might be
more than one correct tweet associated to the same sub-event). Alternatively, tweets are classified
as novel (they contain relevant information for the summary which is not in the reference) or noisy.
From these annotations, we computed the following values for analysis and evaluation: (i) recall,
given by the ratio of sub-events in the reference which are covered by a correct tweet in the
summary; and  (ii) precision, given by the ratio of correct + novel tweets from a whole summary
(note that redundancy is not penalized by any of these measures).

\begin{table}[hbt]
 \begin{center}
  \begin{tabular}{| p{3cm} | l | c | c | c |}
   \cline{3-5}
   \multicolumn{2}{ c |}{} & \textbf{es} & \textbf{en} & \textbf{pt} \\
   \hline
   \hline
   \multirow{2}{*}{\textbf{Goals (54)}} & \textbf{TF} & 0.98 & 0.98 & 0.98 \\
   \cline{2-5}
   & \textbf{KLD} & \textbf{1.00} & \textbf{1.00} & \textbf{1.00} \\
   \hline
   \hline
   \multirow{2}{*}{\textbf{Penalties (2)}} & \textbf{TF} & \textbf{1.00} &
\textbf{0.50} & \textbf{1.00} \\
   \cline{2-5}
   & \textbf{KLD} & \textbf{1.00} & \textbf{0.50} & \textbf{1.00} \\
   \hline
   \hline
   \multirow{2}{*}{\textbf{Red cards (12)}} & \textbf{TF} & 0.75 & 0.75 & 0.83
\\
   \cline{2-5}
   & \textbf{KLD} & \textbf{0.92} & \textbf{0.92} & \textbf{1.00} \\
   \hline
   \hline
   \textbf{Disallowed} & \textbf{TF} & \textbf{0.40}
& \textbf{0.50} & \textbf{0.40} \\
   \cline{2-5}
   \textbf{goals (10)} & \textbf{KLD} & \textbf{0.40} & \textbf{0.50} & 0.30 \\
   \hline
   \hline
   \multirow{2}{*}{\textbf{Game starts (26)}} & \textbf{TF} & 0.73 & 0.74 & 0.79
\\
   \cline{2-5}
   & \textbf{KLD} & \textbf{0.84} & \textbf{0.79} & \textbf{0.83} \\
   \hline
   \hline
   \multirow{2}{*}{\textbf{Game ends (26)}} & \textbf{TF} & \textbf{1.00} &
\textbf{1.00} & \textbf{1.00} \\
   \cline{2-5}
   & \textbf{KLD} & \textbf{1.00} & \textbf{1.00} & \textbf{1.00} \\
   \hline
   \hline
   \textbf{Game stops \&} & \textbf{TF} & 0.62 & \textbf{0.60} & 0.57 \\
   \cline{2-5}
   \textbf{resumptions (63)} & \textbf{KLD} & \textbf{0.68} & \textbf{0.60} &
\textbf{0.59} \\
   \hline
   \hline
   \multirow{2}{*}{\textbf{Overall}} & \textbf{TF} & 0.79 & 0.74 & 0.78 \\
   \cline{2-5}
   & \textbf{KLD} & \textbf{0.84} & \textbf{0.77} & \textbf{0.82} \\
   \hline
  \end{tabular}
  \caption{Recall of reported sub-events for summaries in Spanish (es), English (en), and Portuguese
(pt).}
  \label{tab:detection-of-sub-events}
 \end{center}
\end{table}

Table~\ref{tab:detection-of-sub-events} shows recall values as the coverage of
the two approaches over each type of sub-event, as well as the macro-averaged
overall values. These results corroborate that simple state-of-the-art
approaches like TF and KLD score outstanding recall values. Nevertheless, KLD
shows to be slightly superior than TF for recall. Regarding the averages of all
kinds of sub-events, recall values are near or above 80\% for all the languages.
It can also be seen that some sub-events are much easier to detect than others.
It is important that summaries do not miss the fundamental sub-events. For
instance, all the summaries successfully reported all the goals and all the game
ends, which are probably the most emotional moments, when users extremely
coincide sharing. However, other sub-events like game stops and resumptions, or
disallowed goals, were sometimes missed by the summaries, with recall values
near 50\%. This shows that some of these sub-events may not be that shocking
sometimes, depending on the game, so fewer users share about them, and therefore
are harder to find by the summarization system. For instance, one could expect
that users would not express high emotion when a boring game with no goals stops
for half time. Likewise, this shows that those sub-events are less relevant for
the community. In fact, from these summaries, users would perfectly know when a
goal is scored, when it finished, and what is the final result.

\begin{table}[hbt]
 \begin{center}
  \begin{tabular}{| l | c | c | c |}
   \cline{2-4}
   \multicolumn{1}{ l |}{}& \textbf{es} & \textbf{en} & \textbf{pt} \\
   \hline
   \hline
   \textbf{TF} & 0.79 & 0.74 & 0.79 \\
   \hline
   \textbf{KLD} & \textbf{0.84} & \textbf{0.79} & \textbf{0.83} \\
   \hline
  \end{tabular}
  \caption{Precision of summaries in Spanish (es), English (en), and Portuguese
(pt).}
  \label{tab:tf-vs-kld-usefulness}
 \end{center}
\end{table}

Table \ref{tab:tf-vs-kld-usefulness} shows precision values as the ratio of
useful tweets for the three summaries generated in Spanish, English and
Portuguese. The results show that a simple TF approach is relatively good for
the selection of a representative tweet, with precision values above 70\% for
all three languages. As for recall values, KLD does better than TF, with
precision values near or above 80\%. This shows that taking advantage of the
differences between the current sub-event and tweets shared before considerably
helps in the tweet selection. Note also that English summaries reach 0.79
precision even if the tweet stream is, in that case, an order of magnitude
smaller than their Spanish counterpart, suggesting that the method works well at
very different tweeting rates. 

\section{Related Work}
\label{sec:related-work}

Automatic summarization of events from tweets is still in its infancy as a research field. Some have
tackled the task in an offline mode, after the events were finished. For
instance, Hannon et al.~\cite{hannon2011personalized} present an approach for
the automatic generation of video highlights
for soccer games after they finished. They set a fixed number of sub-events that want to be included
in the highlights, and select that many video fragments with the highest tweeting activity. Others,
such as Petrovi\'c et al.~\cite{petrovic2010streaming}, have shown the
potential of Twitter for the
detection and discovery of events from tweets. While some have studied events after they happened,
there is very little research dealing with the real-time study of events to provide near-immediate
information. Zhao et al.~\cite{zhao2011human} detect sub-events occurred during
NFL games, using an
approach based on the increase of the tweeting activity. We set this approach as the baseline in our
sub-event detection process. Afterward, they apply a specific lexicon provided as input to identify
the type of sub-event. Different from this, our approach aims to be independent of the event,
providing a summarized stream instead of categorizing sub-events. Chakrabarti
and Punera~\cite{chakrabarti2011event} were the first to present an approach
--which is based on Hidden Markov
Models-- for constructing real-time summaries of events from tweets. However, their approach
requires prior knowledge of similar events, and so it is not easily applicable to previously unseen
types of events.

\section{Conclusions}
\label{sec:conclusions}

We have presented a two-step summarization approach that, without making use of external knowledge,
identifies relevant sub-events in soccer games and selects a representative tweet for each of them.
Using simple text analysis methods such as KLD, our system generates real-time summaries with
precision and recall values above 80\% when compared to manually built reports. The fact that users
tweet at the same time, with overlapping vocabulary, helps not only detecting that a sub-event
occurs, but also selecting a representative tweet to describe it. Our study also shows that
considering all previous information seen during the event is really helpful to this end, yielding
superior results than taking into account just the most recent activity. The activity for the soccer
games studied in this work varies from 11k to 74k tweets sent, showing that regardless of the
audience tweeting about an event, our method effectively reports the key sub-events occurred during
a game. Finally, all of the most relevant types of sub-events, such as goals and game ends, are
reported almost perfectly. 

Note that our method does not rely on any external knowledge about soccer events (except for the
schedule time to begin), so it can be straightforwardly applied to other kinds of events. As future
work, we intend to evaluate the performance of the method on other kinds of scheduled events such as
award ceremonies, keynote talks, other types of sport events, product presentations, TV shows, etc.

\section{Acknowledgments}

This work has been part-funded by the Education Council of the Regional Government of Madrid, MA2VICMR
(S-2009/TIC-1542), the Innovation project Holopedia (TIN2010-21128-C02-01), the European Community's Seventh
Framework Programme (FP7/ 2007-2013)
under grant agreement nr. 288024 (LiMoSINe pro\-ject) and the
Spanish Ministry of
Education for a doctoral grant (AP2009-0507).

\bibliographystyle{abbrv}
\bibliography{twitter-soccer-arxiv}  

\balancecolumns

\end{document}